\pdfoutput=1
\documentclass[preprint]{elsarticle}
\usepackage{graphicx}  
\usepackage{dcolumn}   
\usepackage{bm}        
\usepackage{amssymb}   
\usepackage{color}     
\usepackage{pifont}
\usepackage{hyperref}

\def\lsim{\mathrel{\rlap{\lower4pt\hbox{\hskip1pt$\sim$}}\raise1pt\hbox{$<$}}}
\def\gsim{\mathrel{\rlap{\lower4pt\hbox{\hskip1pt$\sim$}}\raise1pt\hbox{$>$}}}
\def\MET{{\mbox{$E\kern-0.57em\raise0.19ex\hbox{/}_{T}~$}}}
\def\METnoSpace{{\mbox{$E\kern-0.57em\raise0.19ex\hbox{/}_{T}$}}}

\def\zz{$ZZ$}
\def\grav{$G^{*}$}
\def\mg{$m_{G^{*}}$}
\def\mgthfize{\mbox{$m_{G^{*}}=350$~GeV}}

\def\mgsefize{\mbox{$m_{G^{*}}=750$~GeV}}

\def\kmplnum{\mbox{$k/\bar{m}_{\rm{pl}}=0.1$}}
\def\kmplnumt{\mbox{$(k/\bar{m}_{\rm{pl}})^2$}}

\def\ifb{$\rm{fb}^{-1}$}

\usepackage{atlasphysics} 
\usepackage{lmetdefs} 

\begin{document}


\journal{Physics Letters B}

\title{\vspace{-0.5cm} \flushright{\normalsize{CERN-PH-EP-2012-026}} \\ \vspace{-0.5cm} \flushright{\normalsize{Submitted to Physics Letters B}} \center{Search for new particles decaying to $ZZ$ using final states with leptons and jets with the ATLAS detector in \shat~ proton-proton collisions}}
\date{\today}

\begin{abstract}
A search is presented for a narrow resonance decaying to a pair of $Z$ bosons using data corresponding to $1.02$~\ifb~of integrated luminosity collected by the ATLAS~experiment from $pp$ collisions at \shat. Events containing either four charged leptons (\llll)~or two charged leptons and two jets (\lljj) are analyzed and found to be consistent with the Standard Model background expectation. Lower limits on a resonance mass are set using the Randall-Sundrum (RS1) graviton model as a benchmark. Using both \llll\ and \lljj\ events, an RS1 graviton with \kmplnum\ and mass between 325 and 845~GeV is excluded at~$95\%$~confidence level. In addition, the \llll\ events are used to set a model-independent fiducial cross section limit of $\sigma_{\rm{fid}}(pp \rightarrow X \rightarrow ZZ) < $~0.92~pb~ at 95\% confidence level for any new sources of $ZZ$ production with $m_{ZZ}$ greater than 300 GeV.
\end{abstract}

\maketitle

\section{ Introduction}

The Standard Model (SM) of particle physics allows for resonant production of $Z$ boson pairs (\zz) solely through the production and decay of the Higgs boson. However, some extensions to the SM predict additional mechanisms for resonant \zz~production. For example, models of warped extra dimensions~\cite{rs1,rs2}~predict two such resonances: excited states of the spin-2 graviton (\grav) and the spin-0 radion ($R$). Searches for such gravitons by the ATLAS Collaboration have excluded at 95\% confidence level masses smaller than 1.63 TeV in dilepton final states~\cite{atlasgrav} and smaller than 1.9 TeV in diphoton final states~\cite{atlasgrav_gamma}; CMS has excluded masses below 1.84 TeV in diphoton final states~\cite{cmsgrav_gamma}.  Recent versions of these models~\cite{kkgraviton}~in which all SM fields propagate in these new dimensions predict enhanced coupling of the graviton to the \zz~final state and suppressed decay rates to light fermion and diphoton states. Observation of graviton production and decay to a pair of $Z$~bosons would be striking evidence for physics beyond the Standard Model.

This letter describes the search for a new particle decaying to the \zz~final state using the RS1 excited graviton ($G^{*}$)~as a benchmark model~\cite{rs1}. This search uses $1.02$~\ifb~of integrated luminosity collected between February and June 2011 by the \mbox{ATLAS} detector in \shat~\pp~collisions at the Large Hadron Collider (LHC). Two final states of the \zz~decay are studied. The first, referred to as  \lljj,  where $\ell=e$ or $\mu$, includes events in which one $Z$~boson decays into electrons or muons, and the other $Z$~boson decays into two jets. This channel is also sensitive to di-jet decays of the $W$~boson  in association with a $Z$ boson decaying to a lepton pair. For the second, referred to as  \llll, both $Z$~bosons decay into electrons or muons.  The final state with two pairs of oppositely charged same-flavor leptons, each pair with invariant mass near the $Z$ boson mass, is used to search for anomalous $ZZ$ production.  

Below a graviton mass (\mg) of $500$~GeV, the \llll~channel dominates the combined \llll +\lljj\ sensitivity due to the extremely low background rate. Above $500$~GeV, the background in  \lljj\ yield  decreases rapidly with $m_{G^*}$, and this final state gains importance due to the larger branching fraction. Since no evidence for \grav~$\rightarrow$~\zz~production is found in this analysis, $95\%$~confidence level (CL) limits are presented using the RS1 graviton as a benchmark. Additionally, the simplicity of the \llll\ final state allows for the calculation of fiducial cross section limits which provide a model-independent bound on anomalous \zz~production.

The RS1 graviton has been used as a benchmark in earlier searches for a resonant structure in \zz~final states. The CDF Collaboration used $p\bar{p}$ collisions at $\sqrt{s}=1.96$~TeV with 2.9~\ifb\ of integrated luminosity to exclude such a state with a mass less than 491 GeV~\cite{cdf_zz}~at $95\%$~CL assuming \kmplnum, where $k$~is the curvature scale of the warped extra dimension and $\bar{m}_{\rm{pl}} \equiv m_{\rm{pl}}/\sqrt{8\pi}$~is the reduced Planck mass. A more recent analysis by CDF using 6~\ifb~reports an excess of \llll\ events at high $Z$ boson-pair invariant mass, clustered around $327$~GeV~\cite{cdf_zz2}, although this is not seen in \lljj\ or $\ell\ell\nu\nu$~channels.

\section{Detector}

The ATLAS detector~\cite{atlas:detector}~is a multi-purpose detector with precision tracking, calorimetry and muon spectrometry. The detector covers almost the entire $4\pi$~solid angle surrounding the collision point at the center of a set of subdetectors. Starting at the collision point and moving outwards, the first subdetector reached is the silicon pixel detector followed by the silicon microstrip detector and the transition radiation tracker. These three systems comprise the inner detector (ID) and reconstruct charged particle tracks out to $|\eta|<2.5$\footnote{ATLAS uses a right-handed coordinate system with its origin at the nominal interaction point~(IP) in the center of the detector and the $z$-axis along the beam line. The $x$-axis points from the IP to the center of the LHC ring, and the $y$ axis points upward. Cylindrical coordinates $(R,\phi)$ are used in the transverse plane, $\phi$ being the azimuthal angle around the beam line. The pseudorapidity is defined in terms of the polar angle $\theta$ ($z=r \cos\theta$) as $\eta=-\ln\tan(\theta/2)$. The transverse energy $E_{\rm{T}}$ is defined as $E \sin{\theta}$, where $E$~is the energy associated to the calorimeter cell or energy cluster. Similarly, $p_{\rm{T}}$~is the momentum component transverse to the beam line.}. Particle momentum is measured by the curvature of the tracks as they are deflected in a peak $2$T~magnetic field provided by a solenoid surrounding the ID. The next subsystems reached are the electromagnetic (EM) and hadronic calorimeters. The EM calorimeter is a highly granular liquid argon (LAr) sampling calorimeter with lead absorber plates designed for electron and photon energy measurements. An iron scintillator tile calorimeter provides hadronic energy measurements in the barrel region ($|\eta|<1.7$) while liquid argon with copper absorber plates is used in the endcap and forward regions. Together these detectors allow electromagnetic and hadronic energy measurements out to $|\eta|<4.9$. Behind the calorimeters is the muon spectrometer (MS), which consists of gas-filled chambers and an air-core toroidal magnetic system. This detector measures both the muon momentum and charge out to $|\eta|<2.7$. 

 To trigger readout~\cite{trigger}, full event reconstruction and event storage by the
data acquisition system, electron candidates must have transverse
energy greater than 20 GeV. They must satisfy shower-shape
requirements and correspond to an ID track. Muon candidates must have
transverse momentum greater than 18 GeV and a consistent trajectory
reconstructed in the ID and muon spectrometer. The full trigger chain
uses signals from all muon detectors. These triggers reach their
efficiency plateau at lepton \pT\ thresholds of 20 GeV for muons and
25 GeV for electrons.

\section{Object Reconstruction}

Electrons are reconstructed from energy deposits in the EM calorimeter matched to tracks in the inner detector, and are required to satisfy the `medium' identification requirements described in Ref.~\cite{egamma}. 
Electrons are required to have $E_{\mathrm T}>20 (15)$~GeV in the \lljj(\llll) channel and $|\eta|<1.37$~or \mbox{$1.52<|\eta|<2.47$}.  For tracks with at least four hits in the pixel and silicon strip detectors, the angles  $\eta$ and $\phi$ are defined by the track, otherwise these quantities are computed from the calorimeter cluster position. Finally, all electrons must be isolated from other charged tracks to suppress jets, i.e. the scalar sum of track \pt~for tracks with \pt~$>1$~GeV~surrounding the electron track in a cone of radius $\Delta R=0.2$, where $\Delta R$~is a distance measure in the $\eta$-$\phi$~plane defined as $\sqrt{(\Delta\eta)^{2} + (\Delta\phi)^{2}}$, must be less than $15\%$ ($10\%$)~of the transverse energy of the electron in the $ee\ell\ell$($ee$jj) channel.

Muons are reconstructed from hits in the muon spectrometer~\cite{staco}. The track formed from these hits must match a track found in the ID. The ID track must have a hit in the innermost layer of the pixel detector to reduce backgrounds from heavy-flavour hadron decays.   The muon track is constructed using information from the ID and MS tracks, and the muon \pt, $\eta$, and $\phi$~are defined from the properties of this combined track. Muons are required to have \pt~$>20 (15)$~GeV in the \lljj(\llll) channel. The lower lepton  \pt\ threshold is used for \llll\ to maintain acceptance at low $Z$ boson pair mass; in \lljj\ the background in this low-\pt\ region is very large. Finally, the muon must be isolated from nearby track activity such that the \pt~sum of all tracks surrounding the muon track in a cone of radius $\Delta R =0.2$ is less than $10\%$ ($15\%$)~of the muon track \pt\ in the \lljj(\llll) channel.

For the \lljj\ channel, jets are reconstructed from a collection of three-dimensional topological energy clusters using the anti-$k_{t}$~sequential recombination clustering algorithm~\cite{cacciari-2008} implemented in the FastJet~\cite{Cacciari:2006sm} package with a radius parameter equal to 0.4.  A jet energy scale (JES) correction is applied to account for the energy response and non-uniformity of the EM and hadronic calorimeters~\cite{jetemjes}. Jets are required to have \pt~$>25$~GeV and $|\eta|<2.8$. If an electron and jet overlap within $\Delta R<0.3$, the jet candidate is removed from the event. The missing transverse momentum, \met, is the modulus of the vector sum of transverse energies of topological calorimeter clusters with $|\eta| < 4.5$, corrected for any high quality muons in the event. The \llll\ channel does not consider jets or missing transverse momentum. The \lljj\ channel considers \met\ only for background studies.

All events must have at least one reconstructed vertex with at least three associated tracks with \pt~$>500$~MeV. The vertex with the largest sum of track $p_{\mathrm T}^2$~is defined as the primary interaction vertex. 

To ensure that they originate from the primary vertex, lepton candidates in the \llll\ and $\mu\mu jj$ channels are required to have a longitudinal impact parameter (distance of closest approach) with respect to the primary vertex of less than 10~mm and a transverse impact parameter significance (transverse impact parameter divided by its error) of less than 10. These requirements reduce contamination from both cosmic rays and leptons produced from hadron decays. In the $eejj$ channel this was found to give no improvement in sensitivity.

Scale factors are applied to the simulation to correct for differences in lepton reconstruction and identification efficiencies between simulation and data. These scale factors have values that differ from unity by 0.1\%--2\% for muons~\cite{ATLAS-CONF-2011-063} and 1\%--13\% for electrons depending  on the \pT\ (for muons) or $E_{\rm T}$ (for electrons); the larger corrections seen for electrons affect only the low-$E_{\rm T}$ region, and are due to mis-modeling of lateral shower shapes in simulation~\cite{tandp}. Systematic uncertainties on these scale factors are derived from efficiency measurements in the data. A small smearing is added to the muon \pT\ in the simulation~\cite{ATLAS-CONF-2011-046} so that the $Z\rightarrow \mu\mu$~invariant mass distribution in data is correctly reproduced by the simulation; similarly, small corrections are applied to the calorimeter energy scale and resolution for electrons.

This analysis uses data collected by single and dilepton triggers during the 2011 LHC run at \shat~with \mbox{50 ns~bunch} spacing. The efficiency of these triggers to select signal-like events is $99\pm1\%$. Additionally, only events recorded while all relevant subdetectors were operating properly are used. The total integrated luminosity for all results in this Letter is $1.02\pm0.04$~\ifb~\cite{atlas:lumi,atlas:lumi_2011jul}.

\section{Simulation}

The signal and all backgrounds other than  multi-jet production are modeled using simulated samples created by process-specific Monte Carlo (MC) event generators. Unless otherwise specified the events in these samples are normalized to the product of the production cross section, the final state branching ratio, and the recorded integrated luminosity. The detector response is simulated with \geant4~\cite{atlassim,geant} after which the event is reconstructed. The RS1 \grav~signal events are generated primarily via gluon-gluon fusion with \pythia~6.421~\cite{pythia} using MRST LO*~\cite{mrst} parton distribution functions for \mg~$=325$~and $500-1500$~GeV in $250$~GeV steps. All samples assume the dimensionless coupling parameter \kmplnum.  The model described by \pythia\ generates events which are uniform in cos$\theta^*$, where $\theta^*$ is the angle between the $Z$ boson direction and the beam axis in the graviton rest frame, and does not have enhanced rates of longitudinal $Z$ boson polarization.

Expected backgrounds from diboson production in the SM ($WW, WZ, ZZ$) are modeled using \herwig~and scaled to the next-to-leading order (NLO) production cross sections as computed by \mcfm~6.0 with MRST2007 LO*~\cite{mrst}. \photos~\cite{photos}~is employed to simulate final state photon radiation and \tauola~2.4~\cite{Jadach1993361}~decays all tau leptons. Production of the background processes \wlnu~and \zll~in association with jets is modeled using the \alpgen~\cite{Mangano:2002ea} event generator with CTEQ6L1~\cite{Pumplin:2002vw} interfaced with \herwig~\cite{herwig_long} for parton showering and \jimmy~\cite{jimmy} to model the underlying event. The \sherpa~\cite{sherpa}~event generator is used to cross check the $W$~and $Z$ boson+jets events simulated by \alpgen; the \mcfm~6.0~\cite{mcfm} generator is also used to check the $Z$ boson+jets background estimate.   Both \wlnu~and \zll~samples are scaled to their respective cross sections at next-to-next-to-leading-order (NNLO) in the strong coupling constant, $\alpha_{S}$, as computed with \fewz~2.0~\cite{fewz,Gavin:2010az}. The top pair (\ttbar) and single top-quark ($tb,tqb,tW$) backgrounds are modeled with  the \mcatnlo~3.41~\cite{mcatnlo} generator interfaced with \herwig~and \jimmy. A sample of $\ttbar$~events generated with \powheg~\cite{powheg}~is used to cross check the \mcatnlo~model. Both $\ttbar$~and single top-quark samples are generated assuming a top-quark mass of $172.5$~GeV. The SM cross section for \ttbar~production is known to approximate-NNLO accuracy as computed in Refs.~\cite{Moch:2008qy, Langenfeld:2009tc, Aliev:2010zk}. Single top-quark production cross sections are calculated to next-to-next-to-leading-logarithm order in $\alpha_{S}$~for the $tb$ process~\cite{tb}, and approximate NNLO order for the $tqb$~and $tW$~processes~\cite{tqb_tW}.

In order to describe properly the
effects of multiple proton-proton interactions per bunch crossing, the
Monte Carlo samples contain multiple interactions per beam-crossing,
weighted to match the data. Additional interactions may produce low-energy deposits in the calorimeter, which leads to a systematic uncertainty in the reconstructed jet energy. Lepton identification and reconstruction efficiency is largely unaffected by multiple interactions, due to the use of track-based isolation.   Many of the background models used are data-driven and so naturally account for multiple interactions.

\section{\lljj~Event selection}

Events in the \lljj\ channel must have exactly two isolated electrons or exactly two isolated muons, each with \pt~$>20$~GeV accompanied by two or more jets, each with \pt~$>25$~GeV. The lepton pair mass ($m_{\ell\ell}$) must be consistent with that of a $Z$~boson ($m_{\ell\ell} \in [66,116]$~GeV); the size of this mass window reflects the non-negligible natural width of the $Z$ boson as well as the lepton momemtum resolution. A requirement that the leptons have opposite charge is applied only to dimuon events, where the charge mis-measurement rate is negligible.

Two signal regions are chosen to maximize the sensitivity to a low-mass (\mg~$<500$~GeV) and high-mass (\mg~$\ge 500$~GeV) signal. In the low-mass region, the  \pt\ of the lepton pair system is required to be greater than $50$~GeV, and similarly the system formed by the two highest \pt\ jets is required to have \pt\ greater than $50$~GeV. In the high-mass region, both \pt\ thresholds are raised to $200$~GeV. In both regions, a signal will manifest itself as a peak in the  \lljj\ invariant mass. The signal definition requires that the two jets result from the decay of a $Z$ boson and therefore have an invariant mass near the $Z$ boson pole mass.  The dijet mass, $m_{jj}$, is thus required to be between 65 GeV and 115 GeV for both low- and high-mass signals. This $m_{jj}$ range was chosen to optimize sensitivity.

\subsection{Backgrounds}

The primary background with this event selection is production of a $Z/\gamma^*$~boson  with associated jets. Secondary backgrounds are  $\ttbar$~and diboson production ($WZ,ZZ$). 

Sidebands surrounding the dijet mass window (below 65 GeV and above 115 GeV) are used to normalize the $Z$ boson+jets background separately for the low- and high-mass signal regions. The normalization factor, defined as the ratio of data to $Z$ boson+jets {\sc alpgen} MC prediction, is $93\%$($75\%$)~in the low(high)-mass signal region. These factors agree within $20\%$~ with those obtained from $Z$ boson+jets events simulated with {\sc sherpa} and scaled to the data in the sidebands. 

The uncertainty of the background prediction in the high-mass selection sample is dominated by $Z$ boson+jets background modeling; the main contribution comes from the uncertainty assigned as a relative deviation of the $Z$ boson+jets normalization factor from unity due to limited $m_{jj}$ sideband statistics.  This assigned uncertainty, which leads to an uncertainty of 40\% on the $Z$ boson+jets background normalization, is combined with an additional uncertainty obtained as the difference between the {\sc alpgen} and {\sc sherpa} predictions in the signal region after sideband normalization, leading to a total uncertainty of 43\%. The $Z$ boson+jets background uncertainty in the low-mass selection sample, which amounts to 6\%, is obtained solely from the scale factor differences between the two $m_{jj}$ sidebands.  The $Z$ boson+jets normalization factors are checked by repeating this study with NLO \lljj\ invariant mass distributions in simulated $Z$ boson+jets events generated with \mcfm6.0  and scaled to the data in the sidebands. The JES uncertainty varies between $12-14\%$~for the background estimate and the signal acceptance~\cite{jetemjes}.

The observed event yield in a $t\bar{t}$-dominated region, low-mass sidebands with the additional requirement of \met\ $>80$~GeV, is found to agree with the Monte Carlo prediction. The top-quark pair background uncertainty is determined to be 25\% from a comparison of event yields between \mcatnlo~and \powheg~together with an evaluation of the sensitivity of the background prediction to the amount of initial state and final state radiation. The uncertainty associated with the theoretical production cross section is estimated to be $10\%$~\cite{hathor}. The uncertainty due to lepton energy and \pt~resolution and reconstruction efficiency contribute less than 3\% to the total uncertainty. The trigger selection efficiency and integrated luminosity contribute $1\%$~and $3.7\%$~\cite{atlas:lumi,atlas:lumi_2011jul} relative uncertainties, respectively.

Production of a $W$~boson with associated jets and single top-quark production are found to give rise to negligible backgrounds. A sample of data events with two low-quality electron candidates (which fail at least one of the requirements above) or two non-isolated muon candidates is used to model the shape of the multijet background. The normalization of this background is determined by a fit to the dilepton mass spectrum using the multijet-like sample as one template and the sum of all other Monte Carlo-based backgrounds as the other template. The multijet background within the dilepton mass range ($m_{\ell\ell} \in [66,116]$~GeV) is determined to be less than 1\% (0.1\%) for $eejj$~($\mu\mu jj$) events.

Table~\ref{tab:lljj}~shows the number of events passing the full selection in the data and expected for each background, and for the RS1~graviton with $m_{G*}=350$ and 750 GeV. No additional scale factors are applied to diboson background events. Figure~\ref{fig:lljj_mzz}~shows the predicted and observed $m_{\ell\ell jj}$~distributions for both low- and high-mass signal selections.

\begin{table}[!bt]
\caption{
Expected numbers of $\ell\ell jj$~events in 1.02 fb$^{-1}$ for each background for the low- and high-mass signal selection regions. The predicted signal yields for $350$~and $750$~GeV graviton signals (\kmplnum) and the observed number of events are also shown. Uncertainties are systematic and statistical.}
\label{tab:lljj}
\begin{tabular}{lD{,}{\,\pm\,}{-1}D{,}{\,\pm\,}{-1}}
\hline\hline
Process&\multicolumn{1}{c}{Low-mass Selection}&\multicolumn{1}{r}{High-mass Selection}\\
\hline
        $Z$+jets & 3530,                 190  & 60,                   27  \\ 
             Top & 81,                   25   & 0.4,                  0.3 \\ 
         Diboson & 92,                   14   & 4,                    1   \\ 
        $W$+jets & 9,                    5    & 1,                    1   \\ 
        Multijet & 14,                   14   & 0.2,                  0.2 \\ \hline
  Background Sum & 3720,                 200  & 66,                   27  \\ \hline
Graviton Signal  & & \\
~~~\mgthfize &  680,                 120  &                           \\ 
~~~\mgsefize &                            & 21,                    4  \\ \hline
            Data & 3515                       & 85                        \\ 
\hline\hline
\end{tabular}
\end{table}

\begin{figure}
\begin{center}
\includegraphics[width=0.7\textwidth]{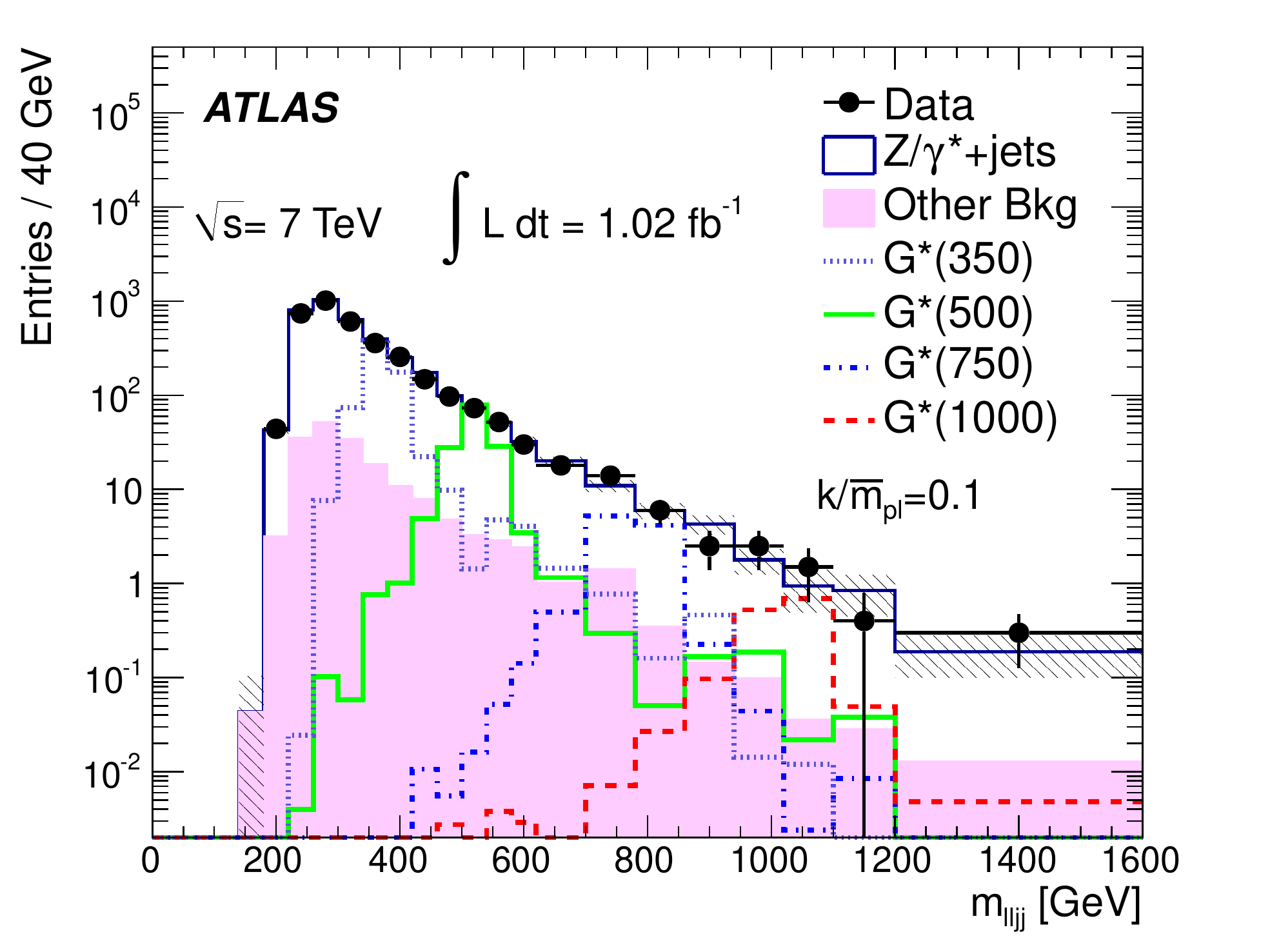}
\includegraphics[width=0.7\textwidth]{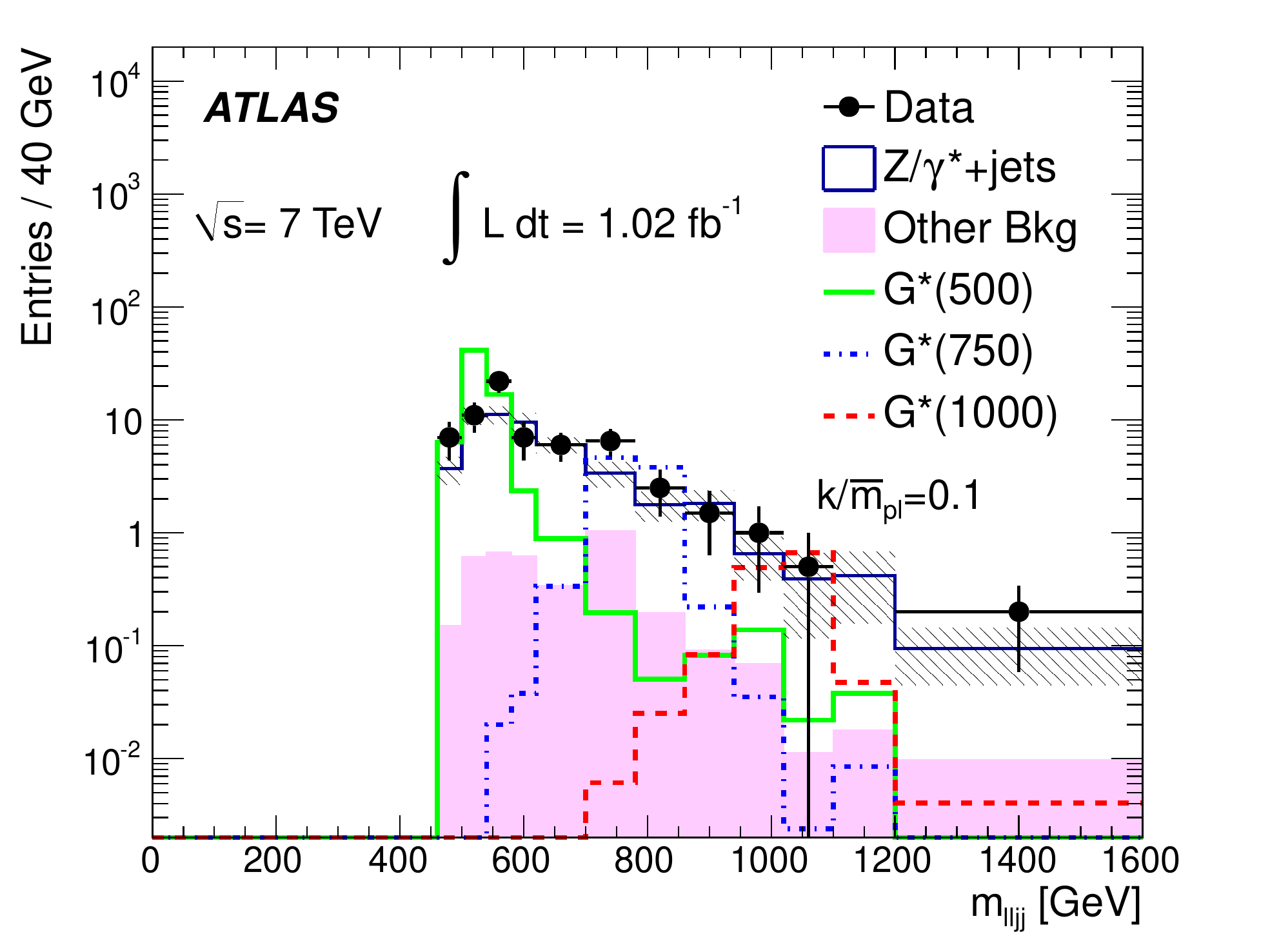}
\end{center}
\caption{Distribution of \lljj\ invariant mass for events satisfying the low-mass signal selection (upper) and high-mass signal selection (lower). These distributions contain both $eejj$~and $\mu\mu jj$~events.  The hatched area shows the uncertainty on the background prediction. }
\label{fig:lljj_mzz}
\end{figure}

\section{\llll\ Event selection}

Events in the \llll\ final state are characterized by at least four high-\pT, isolated electrons
or muons.  Events are required to have passed either a single-muon or single electron trigger which have thresholds of \pT$>18$ GeV and \pT$>20$ GeV, respectively. To minimize the systematic uncertainty on the trigger efficiency, at least one of the selected muons (electrons) is required to have \pT\ $> 20$ ($\ET > 25$) GeV, above which the trigger efficiency dependence on \pT\ ($\ET$) is small. The trigger efficiency for selected events is consistent with 100\% with an uncertainty of 0.04\%.

Same-flavor, oppositely-charged lepton pairs are combined to form $Z$ boson candidates. When more than one such pairing exists, the set with the smallest value of the sum of the two  $|m_{\ell\ell} - \mZ|$ values is chosen. Both $Z$~boson candidates are required to have a dilepton invariant mass  $m_{\ell\ell} \in [66,116]$~GeV; events with two electrons and two muons are categorized as $e^+e^-\mu^+\mu^-$ ($\mu^+\mu^-e^+e^-$) if $m_{e^+e^-}$ ($m_{\mu^+\mu^-}$) is closer to $m_Z$ than $m_{\mu^+\mu^-}$ ($m_{e^+e^-}$). The invariant mass of the \zz\ diboson system must be greater than 300 GeV.   No requirement is made of the \pT~of the individual $Z$ bosons, nor on the \pT~of the $ZZ$ system.

The dominant systematic uncertainties arise from electron identification and muon reconstruction efficiency which range from 3.1\% to 6.6\% and 1.0\% to 2.0\%, respectively, depending on the final state.

\subsection{Backgrounds}

The primary SM source of events with four charged leptons is $(Z/\gamma^*)(Z/\gamma^*)$~production, which we abbreviate as \zz. Other sources are $Z$ (or $W$) boson production in association with additional jets or photons ($W/Z+X$), and top-quark pair
production. The jets might be misidentified as electrons or contain
electrons, photons or muons from in-flight decays of
 pions, kaons, or heavy-flavored hadrons; photons might be misidentified as electrons.
Only a small minority of these background (``misidentified'') leptons survive the isolation requirement. This background is estimated directly from the data.


To estimate the background contribution to the selected sample from events in which one lepton originates from such mis-identified jets, a sample of data events containing three leptons passing all selection criteria plus one `lepton-like jet' is identified; such events are denoted $\ell \ell \ell F$. For muons, the lepton-like jets are muon candidates that fail the isolation requirement. For electrons, the lepton-like jets are clusters in the electromagnetic calorimeter matched to ID tracks that fail either the full electron quality requirements or the isolation requirement or both. The events are otherwise required to pass the full event selection, treating the lepton-like jet as if it were a fully identified lepton. This event sample is dominated by $Z$ boson+jets events. The background is  estimated by scaling the $\ell\ell\ell F$ control sample by a measured factor $f$ (\eta\ and \pT\ dependent and treated as uncorrelated in the two variables) which is the ratio of the probability for a jet to satisfy the full lepton criteria to the probability to satisfy the lepton-like jet criteria. The background in which two selected leptons originate from jets is treated similarly, by identifying a data sample with two leptons and two lepton-like jets; such events are denoted $\ell \ell F F$. To avoid double counting in the background estimate, and to account for the expected \zz~contribution in the control region, $N(ZZ)$, the total number of background events $N(\rm{BG})$ is calculated as:

\begin{equation}
N({\rm BG}) = N(\ell\ell\ell F)\times f - N(\ell\ell FF)\times f^2 - N(ZZ).
\end{equation}

The factor $f$ is measured in a sample of data selected with single-lepton triggers with cuts applied to suppress isolated leptons from $W$ and $Z$ bosons, and corrected for the remaining small contribution of true leptons from $W$ and $Z$ boson decays using simulation.  The negative contribution proportional to $f^2$ is used to correct for double-counting in the term proportional to $f$. A similar analysis is performed on Monte Carlo simulation of background processes of heavy-flavor and light-flavor multi-jet production; the difference between data and simulation is taken as the systematic uncertainty in each \pT\ (or $\eta$) bin. This results in an average systematic uncertainty of $\sim$30\% for each \pT (\eta)  bin except for the lowest \pT\ bin (15-20 \GeV), for which there is nearly a 100\% systematic uncertainty.

In some cases, the control regions from which the background estimate is extrapolated ($\ell\ell\ell F$ or $\ell \ell F F$) contain zero observed events. In such cases, the 68\% CL upper limit on the mean of a Poisson distribution from which zero events are observed is $N<1.29$. We consider the number of events in these regions to be $N = 0.0^{+1.3}_{-0.0}$, and the estimate of the misidentified lepton background uses the value of the lepton misidentification rate $f$ in the lowest \pt\ bin  (15-20 \GeV), which has the largest misidentification rate. This is less likely to happen in electron final states ($e^+e^-e^+e^-$ or $e^+e^-\mu^+\mu^-$), which have two ways for the electron candidate to fail the full selection but still enter the control region, whereas muons are allowed only to fail the isolation requirement. For example, in final states with a muon this leads to $N(\ell\ell\ell F)\times f = 0.0^{+1.3}_{-0.0} \times 0.8 = 0^{+1.0}_{-0.0}$.  When multiple final states are combined, this technique is applied to the combined final state, rather than adding the individual final states in quadrature.  The systematic uncertainty in such cases is evaluated using the misidentification rate uncertainty in the lowest \pT\ bin.

Modeling of the \zz~and non-\zz~SM backgrounds is verified in two data subsamples. To validate the modeling of the \zz~background, events with two opposite-sign same-flavour (OS-SF) pairs, both within a dilepton invariant mass window of $m_{\ell\ell} \in [66,116]$ \GeV, are examined. Requiring two OS-SF pairs inside the chosen $Z$ boson mass window results in an almost pure sample of \zz~events.  To be orthogonal to the signal region for the graviton search, $m_{\ell\ell\ell\ell}<300$ \GeV\ is required. A comparison between the SM \zz~expectation and the observation shows agreement within statistics (see Table~\ref{tab:ctrl1}), indicating satisfactory modeling of the SM \zz~production. 

Requiring four leptons and fewer than two OS-SF pairs but applying the same dilepton mass window used for \zz~pairs to the dilepton pair masses rejects nearly all of the SM \zz~production, so that one may test the misidentified lepton background estimate. This region is orthogonal to the $G^{*}\rightarrow ZZ$ signal regions.  The expected \zz~contribution is $0.15 \pm 0.01 \pm 0.01$, while the misidentified lepton background is $0.0^{+1.3}_{-0.0}\ ^{+0.8}_{-0.0}$. No events are observed in this region, demonstrating an agreement between data and the modeling of misidentified leptons within the available statistics.

\begin{table}
\caption{Expected background contributions (\zz~and misidentified lepton) and
  observed events inside the \zz~control
  region in 1.02 fb$^{-1}$, for events with two opposite-sign
  same-flavor pairs, each with mass $m_{\ell\ell} \in [66,116]$ \GeV, and
  the four lepton mass
  $m_{\ell\ell\ell\ell}<300$ \GeV. The first quoted uncertainty is statistical; the second systematic. }
\label{tab:ctrl1}
\begin{center}
\begin{tabular}{lrrrrrrr}
\hline\hline
Process& $e^+e^-e^+e^-$& $\mu^+\mu^-\mu^+\mu^-$ & $e^+e^-\mu^+\mu^-$\\
& &  & $+\mu^+\mu^-e^+e^-$\\\hline
 \zz~ & $1.3 \pm 0.1 \pm 0.1 $ & $2.5 \pm 0.1\pm 0.1$ & $3.6 \pm 0.1 \pm 0.1$\\
Mis. lep. \vspace{0.1cm}        & $0.01\ _{-0.01}^{+0.02}\ ^{+0.02}_{-0.01}$& $0.3\ ^{+0.9}_{-0.3}\pm
0.2$ &  $0.0\ _{-0.0}^{+1.0}\ ^{+0.8}_{-0.0}$\\ \hline
 Total Bkg.   & $1.3 \pm 0.1 \pm 0.1$& $2.7\ ^{+0.9}_{-0.3}\pm  0.3$ & $3.6\ _{-0.1}^{+1.0}\ ^{+0.8}_{-0.1}$ \\
 Data & 2& 6 & 1\\
\hline\hline
\end{tabular}
\end{center}
\end{table}

Table~\ref{t-signal-yields-llll} shows the expected
 yield in the $m_{\ell\ell\ell\ell}>300$ \GeV\ region. A total of   $1.9^{+1.0}_{-0.1}\
 ^{+0.8}_{-0.1}$ events are expected from SM processes.  Three
 events are observed, see Fig~\ref{fig:llll_mzz}. Due to the asymmetry of the uncertainties,
 three events corresponds to the median expected number of observed
 events from SM processes.

\begin{table}
\caption{Background estimates in 1.02 fb$^{-1}$ of data in the 
  $m_{\ell\ell\ell\ell}>300$\,\GeV\ signal region.  Also shown are expected yields for
  $G^{*}\rightarrow ZZ$ samples for a coupling of \kmplnum.  The first quoted uncertainty is statistical; the second systematic.  }
\label{t-signal-yields-llll}
\begin{center}
\begin{tabular}{lr}
\hline \hline
Process &{\bf Total} \\ \hline
\zz~& $1.9 \pm 0.1\pm 0.1$ \\
Misident. leptons& \vspace{0.1cm} $0.02\ ^{+1.0}_{-0.01}\ ^{+0.8}_{-0.02}$ \\ \hline
Total Bkg. \vspace{0.1cm}  & $1.9\ ^{+1.0}_{-0.1}\ ^{+0.8}_{-0.1}$\\\hline
Data & 3\\
\hline
$G^{*}$(325\,\GeV) & $590 \pm 40 \pm 30$ \\
$G^{*}$(350\,\GeV) & $71 \pm 3 \pm 4$ \\
$G^{*}$(500\,\GeV) & $12 \pm 0.5 \pm 0.6$ \\
$G^{*}$(750\,\GeV) & $1.5 \pm 0.1\pm 0.1$ \\
$G^{*}$(1000\,\GeV) & $(2.7 \pm 0.2 \pm 0.1)\times 10^{-1}$ \\
$G^{*}$(1250\,\GeV) & $(6.6 \pm 0.4 \pm 0.3)\times 10^{-2}$ \\
$G^{*}$(1500\,\GeV) & $(1.9 \pm 0.1 \pm 0.1)\times 10^{-2}$ \\
\hline \hline
\end{tabular}
\end{center}
\end{table}

\begin{figure}
\begin{center}
\includegraphics[width=0.7\textwidth]{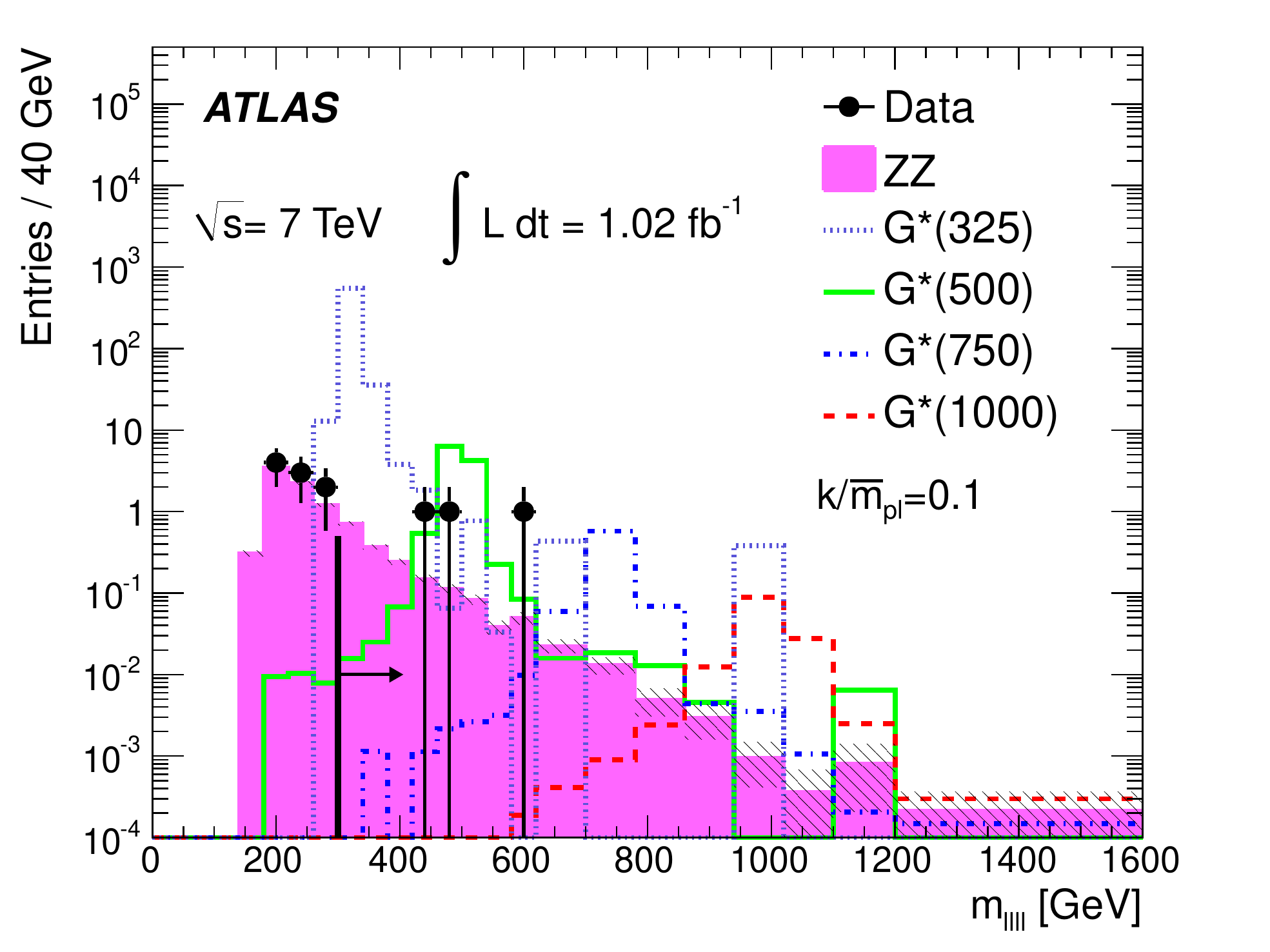}
\end{center}
\caption{Distribution of the four lepton invariant mass for the selected events. The misidentified lepton
  background is negligible and not shown.  Hypothetical graviton signal distributions
  are overlaid. The hatched area shows the uncertainty on the background prediction. The region with $m_{\ell\ell\ell\ell}<300$ \GeV, to the left of
  the solid black line, serves as a
  \zz~control region; the signal region, indicated by the arrow, is $m_{\ell\ell\ell\ell}>300$
  \GeV. Overflow events are shown in the highest mass bin. Numerical
  values are given in Table~\ref{t-signal-yields-llll}.}
\label{fig:llll_mzz}
\end{figure}

\section{Statistical analysis}

To test for possible resonances we search for an excess in the full spectrum using the {\sc bumphunter} algorithm~\cite{bumphunter}.  No significant excess is found in the \llll, low-mass \lljj\ or high-mass \lljj\ spectra. The largest excesses have $p$-values of 0.07, 0.08, and 0.08 respectively, corresponding to significances of $1.5\sigma, 1.4\sigma, 1.4\sigma$, respectively.

Observing no significant excess, we calculate limits on the production cross section times branching ratio for a narrow $ZZ$ resonance from the \llll\ channel and the \lljj\ channels separately, as well as for the combined channel. In the \lljj\ channel, the background falls quickly and the resonance is expected to be fairly narrow; statistical analysis for each hypothesized mass is therefore done as a counting experiment using a single bin that surrounds the hypothesized mass.  The mass windows are chosen to optimize the expected limit in the background-only hypothesis. In the \llll\ channel, the background is very low and the knowledge of the mass dependence of the misidentified lepton background is limited by the small number of events in the sample used to estimate its contribution. Hence, a single wide window, \mbox{$m_{\llll}>300$}~\GeV, is used.  Limits are evaluated at a specific set of mass points and interpolated between them, as the background levels and signal acceptance are smoothly varying.

Limits are set  using the CLs method~\cite{junk,read}, a modified frequentist approach. In this method a log-likelihood ratio (LLR) test statistic is formed using the Poisson probabilities for estimated background yields, the signal acceptance, and the observed number of events for all \zz\ resonance mass hypotheses, accounting for systematic
uncertainties on the background estimate and signal acceptance. Pseudo-experiments are drawn from a Poisson distribution whose mean is drawn from a bifurcated Gaussian
and truncated at zero; a bifurcated Gaussian has distinct positive and negative widths to represent asymmetric uncertainties.  Confidence levels are derived by integrating the LLR~in pseudo-experiments using both the signal plus background hypotheses ($CL_{s+b}$)~as well as the background only hypothesis ($CL_{b}$). In the modified frequentist approach, the  production cross section excluded at 95\% CL is computed as the cross section for which $CL_{s}$, defined as $CL_{s+b}/CL_{b}$, is equal to 0.05.

\subsection{\lljj\ limits}

For the statistical analysis of the \lljj\ data, mass windows are chosen surrounding each of the generated graviton masses to perform a counting experiment, as shown in Table~\ref{tab:lljj_windows}.    The mass windows are chosen to optimize the expected limit in the background-only hypothesis.  For a resonance mass of 350 GeV, the low-mass selection described above is as the initial selection; at the remaining mass values the high-mass selection described above is used. Observed limits are shown in Table~\ref{tab:lims} for the individual $eejj$ and $\mu\mu jj$ channels. The combined limits for the \lljj\ channel are shown in Figure~\ref{fig:lljj_mzz_lim}.

\begin{figure}
\begin{center}
\includegraphics[width=0.7\textwidth]{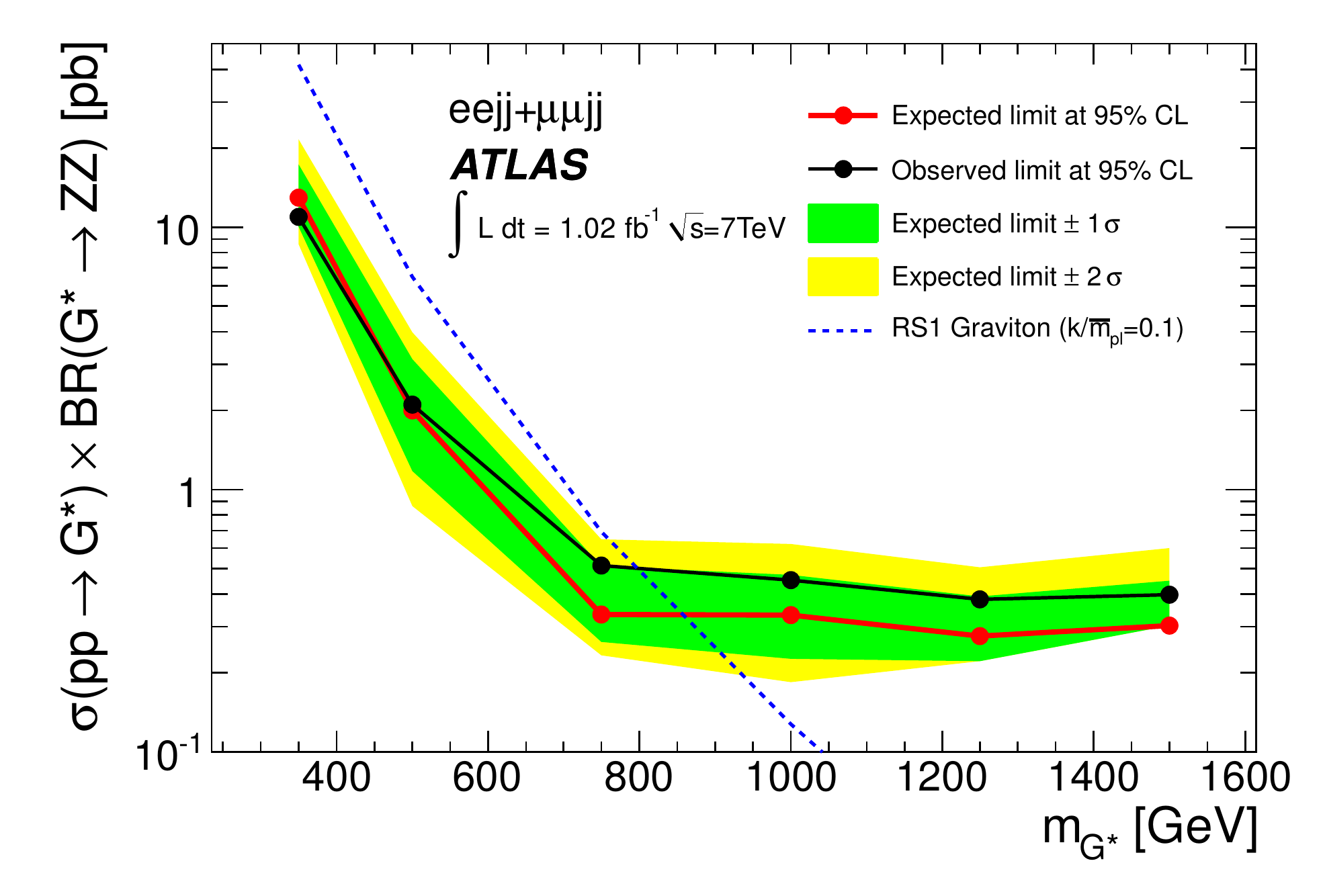}
\end{center}
\caption{Expected and observed  95\% CL limits for $G^{*}\rightarrow ZZ$~for the combined \lljj\ channel. The leading-order theoretical prediction is also shown for \kmplnum.  Theoretical predictions scale with \kmplnumt.}
\label{fig:lljj_mzz_lim}
\end{figure}

The acceptance times efficiency including the mass window cuts is 5.6\% (5.3\%, 13.1\%, 12.9\%, 8.6\%, and 6.0\%) for 350 (500, 750, 1000, 1250 and 1500) GeV graviton signal with $Z$ boson decays to $e^+e^-$, $\mu^+\mu^-$, and $\tau^+\tau^-$; the uncertainty is 15\%, relative. The acceptance drop at high mass is due to highly boosted $Z$ bosons producing a single merged jet.

\begin{table}
\caption{Mass windows (in GeV) for statistical analysis of the \lljj\ channel at each of the generated graviton masses (in GeV), with the expected numbers of background, graviton (\kmplnum), and observed events. For a resonance mass of 350 GeV, the low-mass selection is used; at the remaining mass values the high-mass selection is used. Uncertainties are statistical and systematic added in quadrature.}
\label{tab:lljj_windows}
\begin{tabular}{lllrrr}
\hline\hline
Resonance & Mass & & Expected & Expected & \\
Mass  & Window  & &Background & Signal & Obs. \\ \hline\noalign{\vspace{1mm}}
350  & $330-360$      & $eejj$      & $116^{+20}_{-15}$   & $161^{+36}_{-14}$    & 109 \\[1mm]
     &                & $\mu\mu jj$ & $163^{+28}_{-23}$   & $165^{+19}_{-16}$    & 147 \\[1mm]
500  & $480-530$      & $eejj$      & $6^{+4}_{-2}$       & $27^{+3}_{-4}$       & 8   \\[1mm]
     &                & $\mu\mu jj$ & $8^{+5}_{-2}$       & $23^{+2}_{-3}$       & 6   \\[1mm]
750  & $730-830$      & $eejj$      & $4^{+2}_{-1}$       & $6.5^{+0.6}_{-0.9}$  & 6   \\[1mm]
     &                & $\mu\mu jj$ & $1.2^{+0.9}_{-0.5}$ & $6.9^{+0.6}_{-0.7}$  & 2   \\[1mm]
1000 & $900-1090$     & $eejj$      & $2.1^{+1.3}_{-0.9}$ & $1.2\pm 0.2$  & 2   \\[1mm]
     &                & $\mu\mu jj$ & $1.2^{+0.8}_{-0.5}$ & $1.2\pm 0.1$  & 3   \\[1mm]
1250 & $\ge 1150$  & $eejj$      & $0.4^{+0.4}_{-0.3}$ & $0.18\pm 0.01$ & 1   \\[1mm]
     &                & $\mu\mu jj$ & $0.5^{+0.5}_{-0.4}$ & $0.21\pm 0.01$ & 1   \\[1mm]
1500 & $\ge 1300$ & $eejj$      & $0.1\pm 0.1$ & $0.04\pm 0.01$ & 0   \\[1mm]
     &                & $\mu\mu jj$ & $0.4\pm 0.4$ & $0.04\pm 0.01$ & 1   \\[1mm]
\hline\hline
\end{tabular}\end{table}

\subsection{\llll\ limits}

The analysis of the \llll\ data is done using a single mass-independent counting experiment. The median expected upper limit on the number
of \llll\ events from a new source with \mbox{$m_{\llll}>300$}~\GeV~is $N_{ZZ}<5.7$ events at 95\% CL. The observed three
events leads to an upper limit of $N_{ZZ}<5.7$ events at 95\% CL.

We define a $ZZ\rightarrow\ell\ell\ell\ell$  fiducial region, which contains
events with four charged leptons ($e$ or $\mu$) each with $\pt >15$ \GeV\ and $|\eta |<2.5$ forming two OS-SF pairs each with $m_{\ell\ell} \in
[66,116]$ \GeV\ and  \mbox{$m_{\llll}>300$}~\GeV.  Within this fiducial region, the efficiency of the
\llll\ selection is nearly independent of the graviton mass for the RS1
graviton benchmark model, as shown in Table~\ref{tab:grav_pvals}.

 The lowest selection efficiency (61\%) is used
  to set limits  on all mass points. The corresponding
\zz~fiducial limit on the production of new sources of high-mass
\zz~pairs is

\[ \sigma_{ ZZ{\rm,\ fid}} < \frac{  N_{ZZ}  }{ \epsilon_{ZZ}\times
  {\rm B}(ZZ\rightarrow \ell\ell\ell\ell) \times \mathcal{L} } \]
\[ = \frac{ 5.7
}{ 0.61 \times 0.010 \times 1.02\ {\rm
    fb}^{-1}} =  0.92\ {\rm pb},\]

\noindent
which can be applied to our benchmark model of RS1 gravitons using the
fiducial acceptance, see Figure~\ref{fig:llll_mzz_lim} and
Table~\ref{tab:grav_pvals}, but may be extended to a larger class of
models that hypothesize resonances with branching fraction (B) to $ZZ$
different than that in the RS1 model.  

 The fiducial efficiency is relative to all $ZZ$ decays to charged leptons, including $\tau$ leptons, and therefore B$(ZZ\rightarrow \ell\ell\ell\ell) = 0.010$ also includes $\tau$ lepton decays.

\begin{figure}
\begin{center}
\includegraphics[width=0.7\textwidth]{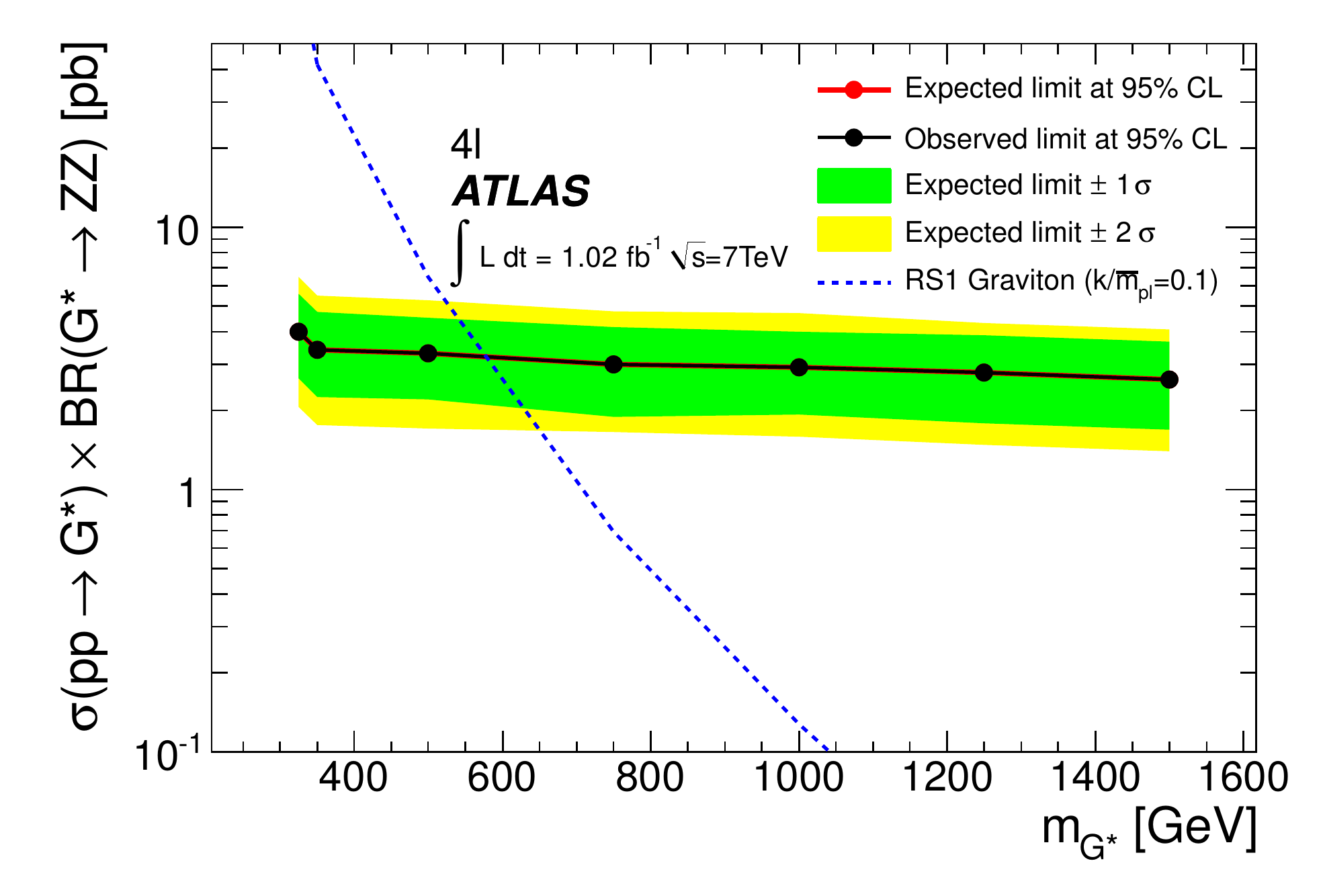}
\end{center}
\caption{Expected and observed  95\% CL limits for $G^{*}\rightarrow ZZ$
using
  the \zz$\rightarrow \ell\ell\ell\ell$~fiducial cross section upper limit, and the fraction of
  the produced events which fall into the fiducial region, see
  Table~\ref{tab:grav_pvals}.  The median expected result from
  pseudo-experiments drawn from the SM hypothesis is shown; the
  observed limit corresponds to $N_{obs}=3$.
 The leading-order theoretical prediction is also shown for \kmplnum. Theoretical predictions scale with \kmplnumt.}
\label{fig:llll_mzz_lim}
\end{figure}

\begin{table}
\caption{For spin-2 RS1 graviton 
  models, the theoretical prediction of $\sigma(pp\rightarrow
  G^{*}\rightarrow ZZ)$ using  a coupling of
  \kmplnum; acceptance of the $G^{*}\rightarrow ZZ\rightarrow \ell\ell\ell\ell$ fiducial
  region defined in the text, and the efficiency of the selection described in
  the text; the median expected and observed 95\% CL upper limits on
  $\sigma(pp\rightarrow
  G^{*}\rightarrow ZZ)$ using the fiducial cross-section limit.  We use
  B($G^{*}\rightarrow ZZ$)=4.5\% ~\cite{gbr} and $\sigma(pp\rightarrow
  G^{*})$ at leading order from {\sc pythia} with \kmplnum; predictions  scale as \kmplnumt. As is done with the \llll\ fiducial limit, the lowest selection efficiency (61\%) is used
  to set limits  on all mass points.
}
\label{tab:grav_pvals}
\begin{center}
\begin{tabular}{lrrrrr}
\hline\hline
Graviton& Theory   & Fid. & Sel. &  Exp.  & Obs.
\\ 
 Mass  & {\scriptsize $k/\bar{m}_{\textrm{pl}}$ = 0.1}  & Acc. & Eff. &  Limit &  Limit  \\ 
 (GeV) & (pb)  & & & (pb) & (pb) \\ \hline
325   & 950   & 23\% & 61\% & 4.0    & 4.0\\
350   & 42  & 27\% & 61\% & 3.3    & 3.3\\
500   & 6.5    & 28\% & 63\% & 3.2    & 3.2\\
750   & 0.69    & 31\% & 66\% & 2.9    & 2.9\\
1000  & 0.13   & 32\% & 66\% & 2.8    & 2.8\\
1250  & 0.03    & 33\% & 67\% & 2.7    & 2.7\\
1500  & 0.01    & 35\% & 66\% & 2.6    & 2.6\\
\hline\hline
\end{tabular}
\end{center}
\end{table}

\subsection{Combined limits}

The limits obtained from combinations of channels are calculated using the same technique, keeping each channel separate but with a coherent signal hypothesis and including the correlations between the systematics where appropriate. The dominant uncertanties in the \llll\ channel are due to the misidentified-lepton estimate; the degree of correlation with uncertainties in the \lljj\ channel is small.  The combined limits are shown in Fig~\ref{fig:comb_mzz_lim} and given in  Table~\ref{tab:lims}.

\begin{figure}
\begin{center}
\includegraphics[width=0.7\textwidth]{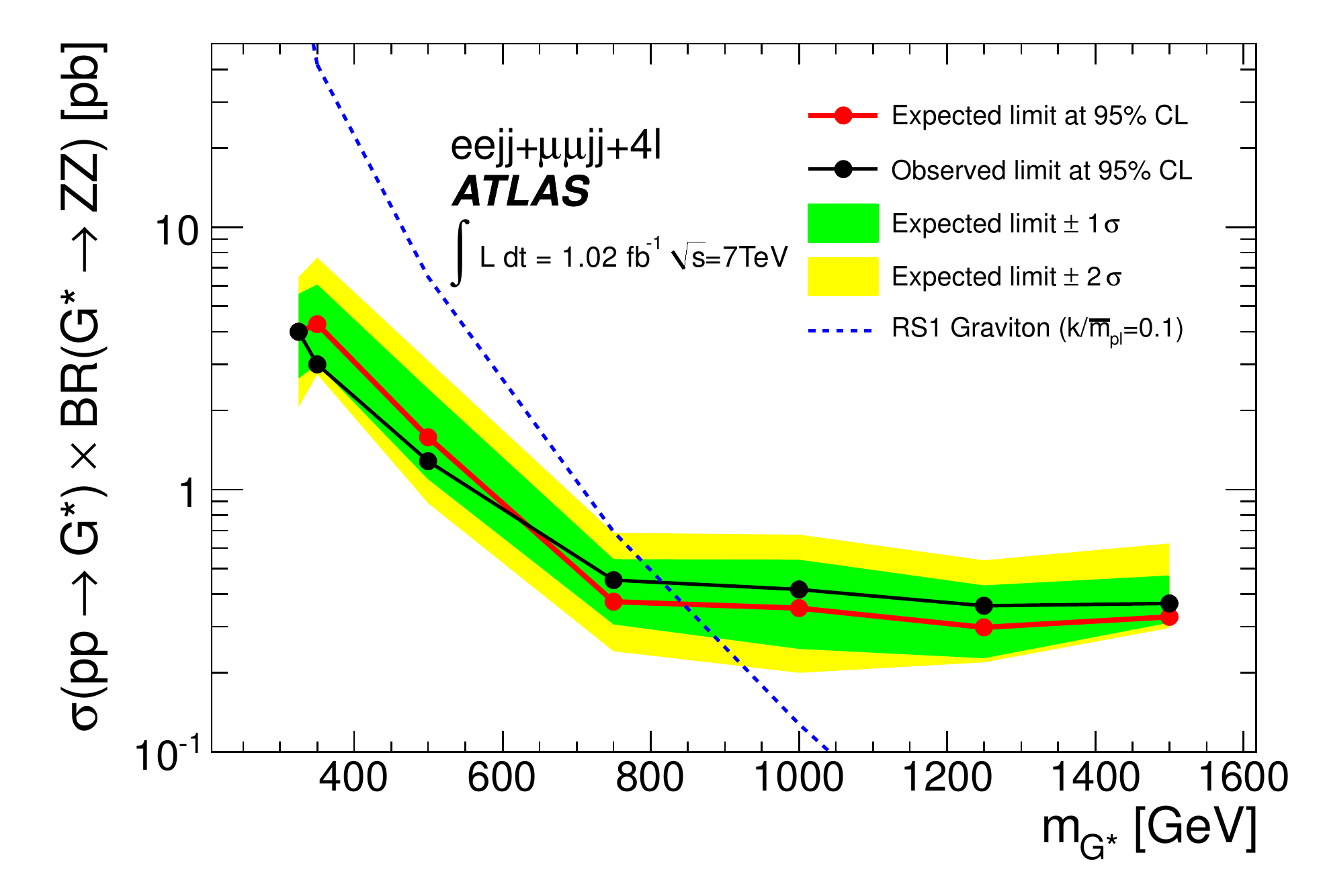}
\end{center}
\caption{Expected and observed  95\% CL limits for $G^{*}\rightarrow ZZ$~for the combined  \lljj+\llll\  channels. The leading-order theoretical prediction is also shown for \kmplnum.  Theoretical predictions scale with \kmplnumt.}
\label{fig:comb_mzz_lim}
\end{figure}

\begin{table}
\caption{ Upper limits at 95\% CL on $\sigma(pp\rightarrow G^{*}) \times \rm{B}(G^{*}\rightarrow ZZ)$ in the individual channels $eejj$, $\mu\mu jj$ and \llll\ as well as the combined \lljj\ ($ eejj$ and $\mu\mu jj$) and \llll+\lljj\ ($ eejj$, $\mu\mu jj$ and \llll) channels. }
\label{tab:lims}
\begin{tabular}{lrrrrrr}
\hline\hline
Graviton Mass    &  $eejj$\ \  & $\mu\mu jj$\ \  & \lljj\ \  & \llll\ \  & \llll+\lljj \\ 
(GeV) & (pb) & (pb) & (pb) & (pb) & (pb) \\ \hline
325      & -- & -- & -- & 4.0 & 4.0 \\
350      & 8.9   & 11.6  & 10.9  & 3.3   & 3.0   & \\
500      & 2.3   & 1.8   & 2.1   & 3.3   & 1.3   & \\
750      & 0.9   & 0.5   & 0.5   & 2.9   & 0.5   & \\
1000     & 0.6   & 0.7   & 0.5   & 2.8   & 0.4   & \\
1250     & 0.7   & 0.6   & 0.4   & 2.8   & 0.4   & \\
1500     & 0.7   & 0.9   & 0.4   & 2.6   & 0.4   & \\
\hline\hline
\end{tabular}\end{table}

We exclude an RS1 graviton in the mass range of 325 to 845~GeV at $95\%$~CL for \kmplnum.

\section{Conclusions}

We report the results of a search for narrow resonances such as Randall-Sundrum gravitons decaying to $ZZ$ pairs, using \lljj\ and \llll\ final states collected by the ATLAS detector from \shat~LHC \pp~collisions. No excess is seen above the expected SM backgrounds, and upper limits are set on the cross section of graviton production times the branching fraction to $ZZ$. We exclude an RS1 graviton in the mass range of 325 to 845~GeV at $95\%$~CL for \kmplnum.
   
\global\emergencystretch = .2\hsize

\section*{Acknowledgments}

We thank CERN for the very successful operation of the LHC, as well as the
support staff from our institutions without whom ATLAS could not be
operated efficiently.

We acknowledge the support of ANPCyT, Argentina; YerPhI, Armenia; ARC,
Australia; BMWF, Austria; ANAS, Azerbaijan; SSTC, Belarus; CNPq and FAPESP,
Brazil; NSERC, NRC and CFI, Canada; CERN; CONICYT, Chile; CAS, MOST and NSFC,
China; COLCIENCIAS, Colombia; MSMT CR, MPO CR and VSC CR, Czech Republic;
DNRF, DNSRC and Lundbeck Foundation, Denmark; EPLANET and ERC, European Union;
IN2P3-CNRS, CEA-DSM/IRFU, France; GNAS, Georgia; BMBF, DFG, HGF, MPG and AvH
Foundation, Germany; GSRT, Greece; ISF, MINERVA, GIF, DIP and Benoziyo Center,
Israel; INFN, Italy; MEXT and JSPS, Japan; CNRST, Morocco; FOM and NWO,
Netherlands; RCN, Norway; MNiSW, Poland; GRICES and FCT, Portugal; MERYS
(MECTS), Romania; MES of Russia and ROSATOM, Russian Federation; JINR; MSTD,
Serbia; MSSR, Slovakia; ARRS and MVZT, Slovenia; DST/NRF, South Africa;
MICINN, Spain; SRC and Wallenberg Foundation, Sweden; SER, SNSF and Cantons of
Bern and Geneva, Switzerland; NSC, Taiwan; TAEK, Turkey; STFC, the Royal
Society and Leverhulme Trust, United Kingdom; DOE and NSF, United States of
America.

The crucial computing support from all WLCG partners is acknowledged
gratefully, in particular from CERN and the ATLAS Tier-1 facilities at
TRIUMF (Canada), NDGF (Denmark, Norway, Sweden), CC-IN2P3 (France),
KIT/GridKA (Germany), INFN-CNAF (Italy), NL-T1 (Netherlands), PIC (Spain),
ASGC (Taiwan), RAL (UK) and BNL (USA) and in the Tier-2 facilities
worldwide.

\newpage


\bibliographystyle{model1-num-names}
\bibliography{letter}{}

\onecolumn
\begin{flushleft}
{\Large The ATLAS Collaboration}

\bigskip

G.~Aad$^{\rm 48}$,
B.~Abbott$^{\rm 110}$,
J.~Abdallah$^{\rm 11}$,
A.A.~Abdelalim$^{\rm 49}$,
A.~Abdesselam$^{\rm 117}$,
O.~Abdinov$^{\rm 10}$,
B.~Abi$^{\rm 111}$,
M.~Abolins$^{\rm 87}$,
O.S.~AbouZeid$^{\rm 157}$,
H.~Abramowicz$^{\rm 152}$
% ...
\bigskip

% ...
$^{10}$ Institute of Physics, Azerbaijan Academy of Sciences, Baku, Azerbaijan\\
$^{11}$ Institut de F\'isica d'Altes Energies and Departament de F\'isica de la Universitat Aut\`onoma  de Barcelona and ICREA, Barcelona, Spain\\
% ...
$^{48}$ Fakult\"{a}t f\"{u}r Mathematik und Physik, Albert-Ludwigs-Universit\"{a}t, Freiburg i.Br., Germany\\
$^{49}$ Section de Physique, Universit\'e de Gen\`eve, Geneva, Switzerland\\
% ...
$^{87}$ Department of Physics and Astronomy, Michigan State University, East Lansing MI, United States of America\\
% ...
$^{110}$ Homer L. Dodge Department of Physics and Astronomy, University of Oklahoma, Norman OK, United States of America\\
$^{111}$ Department of Physics, Oklahoma State University, Stillwater OK, United States of America\\
% ...
$^{117}$ Department of Physics, Oxford University, Oxford, United Kingdom\\
% ...
$^{152}$ Raymond and Beverly Sackler School of Physics and Astronomy, Tel Aviv University, Tel Aviv, Israel\\
% ...
$^{157}$ Department of Physics, University of Toronto, Toronto ON, Canada\\
% ...
\end{flushleft}


\clearpage

\end{document}